\pdfoutput=1
\documentclass[preprint,12pt]{elsarticle}

\usepackage{xspace,graphicx,mparhack,amsmath}
\usepackage{amssymb,url,underscore,fancyvrb,cancel}
\usepackage{hepnicenames,hepunits}
\usepackage{picinpar,fancybox}
\usepackage{microtype,relsize}

\usepackage{rivetversion}

\DeclareFontShape{OT1}{cmtt}{bx}{n}{<5><6><7><8><9><10><10.95><12><14.4><17.28><20.74><24.88>cmttb10}{}

\makeatletter
\g@addto@macro\bfseries{\boldmath}
\makeatother

\newenvironment{snippet}{\Verbatim}{\endVerbatim}

\newcommand{\kbd}[1]{\texttt{#1}\xspace}
\newcommand{\inp}[1]{\textsf{\textdollar}\hspace{1mm}\texttt{#1}\xspace}
\newcommand{\outp}[1]{\textsf{#1}\xspace}
\newcommand{\code}[1]{\texttt{#1}\xspace}
\newcommand{\var}[1]{\texttt{\textdollar{}#1}\xspace}
\newcommand{\val}[1]{\textit{\ensuremath{\langle\text{\textrm{#1}\/}\rangle}}\xspace}
\newcommand{\home}{\texttt{\ensuremath{\sim}}\xspace}

\newcommand{\Delphi}{\textsc{Delphi}\xspace}

\newcommand{\cmdbreak}{\textbackslash\newline}

\newcommand{\AppendixRef}[1]{appendix~\ref{#1}}
\newcommand{\SectionRef}[1]{section~\ref{#1}}

\let\oldmarginpar\marginpar
\renewcommand\marginpar[1]{\-\oldmarginpar{\footnotesize \textit{#1}}}

\newcommand{\warnimg}{\includegraphics[height=11mm]{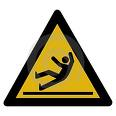}}
\newcommand{\coneimg}{\includegraphics[height=11mm]{cone}}
\newcommand{\bendimg}{\includegraphics[height=11mm]{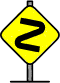}}
\newcommand{\dblbendimg}{\bendimg\hspace{0.5mm}\bendimg}
\newcommand{\thinkimg}{\includegraphics[height=16mm]{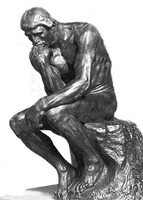}}

\newenvironment{detail}{\begin{window}[0,l,{\bendimg},{}]}{\end{window}\ignorespacesafterend}

\newenvironment{warning}{\vspace{5mm}\hrule\nobreak\vspace{3mm}\begingroup%
  \begin{window}[0,l,{\warnimg},{}]
  \setlength{\parindent}{0cm}\noindent}{%
  \end{window}\endgroup\vspace{3mm}\nobreak\hrule\vspace{5mm}\ignorespacesafterend}

\newenvironment{philosophy}{\vspace{5mm}\hrule\nobreak\vspace{3mm}\begingroup%
  \it\begin{window}[0,l,{\thinkimg},{}]
  \setlength{\parindent}{0cm}\noindent}{%
  \end{window}\endgroup\vspace{3mm}\nobreak\hrule\vspace{5mm}\ignorespacesafterend}

\newcommand{\pT}{\ensuremath{p_\perp}\xspace}

\usepackage{hyperref}

\begin{document}

\begin{frontmatter}
  
  \title{Rivet user manual}

  \author[a]{Andy Buckley}
  \author[b]{Jonathan Butterworth}
  \author[c]{David Grellscheid}
  \author[c]{Hendrik Hoeth}
  \author[d]{Leif L\"onnblad}
  \author[b]{James Monk}
  \author[e]{Holger Schulz}
  \author[f]{Frank Siegert\corref{author}}

  \address[a]{PPE Group, School of Physics, University of Edinburgh, UK.}
  \address[b]{HEP Group, Dept. of Physics and Astronomy, UCL, London, UK.}
  \address[c]{IPPP, Durham University, UK.}
  \address[d]{Theoretical Physics, Lund University, Sweden.}
  \address[e]{Institut f\"ur Physik, Berlin Humboldt University, Germany.}
  \address[f]{Physikalisches Institut, Freiburg University, Germany.}

  \cortext[author]{Corresponding author.\\\textit{E-mail address:} frank.siegert@cern.ch}

  \begin{abstract}
    This is the manual and user guide for the Rivet system for the
    validation and tuning of Monte Carlo event generators. As well as the core
    Rivet library, this manual describes the usage of the \kbd{rivet} program and
    the AGILe generator interface library. The depth and level of description is
    chosen for users of the system, starting with the basics of using validation
    code written by others, and then covering sufficient details to write new
    Rivet analyses and calculational components.
  \end{abstract}


\end{frontmatter}

{\bf PROGRAM SUMMARY}

\begin{small}
  \noindent
  {\em Manuscript Title: Rivet user manual}\\
  {\em Authors: Andy Buckley, Jonathan Butterworth, David Grellscheid,
  Hendrik Hoeth, Leif L\"onnblad, James Monk, Holger Schulz, Frank Siegert}\\
  {\em Program Title: Rivet}\\
  {\em Journal Reference:}                                      \\
  {\em Catalogue identifier:}                                   \\
  {\em Licensing provisions:}                                   \\
  {\em Programming language: C++, Python}\\
  {\em Computer: PC running Linux, Mac}\\
  {\em Operating system: Linux, Mac OS}\\
  {\em RAM: 20M} bytes\\
  {\em Number of processors used: 1}                              \\
  {\em Supplementary material:}                                 
  {\em Classification: 11.9 Event Reconstruction and Data Analysis}\\
  {\em External routines/libraries: HepMC, GSL, FastJet, Python, Swig, Boost, YAML}\\
  {\em Nature of problem:}\\
  Experimental measurements from high-energy particle colliders should be
  defined and stored in a general framework such that it is simple to compare
  theory predictions to them. Rivet is such a framework, and contains at the
  same time a large collection of existing measurements.
  \\
  {\em Solution method:}\\
  Rivet is based on HepMC events, a standardised output format provided by many
  theory simulation tools. Events are processed by Rivet to generate histograms
  for the requested list of analyses, incorporating all experimental phase
  space cuts and histogram definitions.
  \\
  {\em Restrictions:}\\
  Can not calculate statistical errors for correlated events as they appear
  in NLO calculations.
  \\
  {\em Unusual features:}\\
  It is possible for the user to implement and use their own custom analysis
  as a module without having to modify the main Rivet code/installation.
  \\
  {\em Additional comments:}\\
  \\
  {\em Running time:}\\
  Depends on the number and complexity of analyses being applied, but typically
  a few hundred events per second.
  \\

\end{small}

\section{Introduction}
\label{sec:intro}
This manual is a users' guide to using the Rivet generator validation
system. Rivet is a C++ class library, which provides the infrastructure and
calculational tools for particle-level analyses for high energy collider
experiments, enabling physicists to
validate event generator models and tunings with minimal effort and maximum
portability. Rivet is designed to scale effectively to large numbers of analyses
for truly global validation, by transparent use of an automated result caching
system.

The Rivet ethos, if it may be expressed succinctly, is that user analysis code
should be extremely clean and easy to write --- ideally it should be
sufficiently self-explanatory to in itself be a reference to the experimental
analysis algorithm --- without sacrificing power or extensibility. The machinery
to make this possible is intentionally hidden from the view of all but the most
prying users. Generator independence is explicitly required by virtue of all
analyses operating on the generic ``HepMC'' event record.

The simplest way to use Rivet is via the \kbd{rivet} command line tool, which
analyses textual HepMC event records as they are generated and produces output
distributions in a structured textual format. The input events are generated
using the generator's own steering program, if one is provided; for generators
which provide no default way to produce HepMC output, the AGILe generator
interface library, and in particular the \kbd{agile-runmc} command which it
provides, may be useful. For those who wish to embed their analyses in some
larger framework, Rivet can also be used as a library to run programmatically
on HepMC event objects with no special executable being required.

Before we get started, a declaration of intent: this manual is intended to be a
guide to using Rivet, rather than a comprehensive and painstakingly maintained
reference to the application programming interface (API) of the Rivet
library. For that purpose the online documentation at
\url{http://rivet.hepforge.org} should be sufficient -- in case of confusion
please contact the authors at \url{rivet@projects.hepforge.org}. Similar API
documentation is maintained for AGILe at \url{http://agile.hepforge.org}.

\subsection{Typographic conventions}
As is normal in computer user manuals, the typography in this manual is used to
indicate whether we are describing source code elements, commands to be run in a
terminal, the output of a command etc.

The main such clue will be the use of \kbd{typewriter-style} text: this
indicates the name of a command or code element --- class names, function names
etc. Typewriter font is also used for commands to be run in a terminal, but in
this case it will be prefixed by a dollar sign, as in \inp{echo "Hello" |
  cat}.  The output of such a command on the terminal will be typeset in
\outp{sans-serif} font. When we are documenting a code feature in detail (which
is not the main point of this manual), we will use square brackets to indicate
optional arguments, and italic font between angle brackets to represent an
argument name which should be replaced by a value,
e.g. \code{Event::applyProjection(\val{proj})}.


\cleardoublepage

\part{Getting started with Rivet}
\label{part:gettingstarted}
As with many things, Rivet may be meaningfully approached at several distinct
levels of detail:

\begin{itemize}
\item The simplest, and we hope the most common, is to use the analyses which
  are already in the library to study events from a variety of generators and
  tunes: this is enormously valuable in itself and we encourage all manner of
  experimentalists and phenomenologists alike to use Rivet in this mode.
\item A more involved level of usage is to write your own Rivet analyses ---
  this may be done without affecting the installed standard analyses by use of a
  ``plugin'' system (although we encourage users who develop analyses to submit
  them to the Rivet developers for inclusion into a future release of the main
  package). This approach requires some understanding of programming within
  Rivet but you don't \emph{need} to know about exactly what the system is doing
  with the objects that you have defined.
\item Finally, Rivet developers and people who want to do non-standard things
  with their analyses will need to know something about the messy details of
  what Rivet's infrastructure is doing behind the scenes. But you'd probably
  rather be doing some physics!
\end{itemize}

The current part of this manual is for the first sort of user, who wants to get
on with studying some observables with a generator or tune, or comparing several
such models. Since everyone will fall into this category at some point, our
present interest is to get you to that all-important ``physics plots'' stage as
quickly as possible. Analysis authors and Rivet service-mechanics will find the
more detailed information that they crave in Part~\ref{part:writinganalyses}.

\section{Quickstart}

The point of this section is to get you up and running with Rivet as soon as
possible. Doing this by hand may be rather frustrating, as Rivet depends on
several external libraries --- you'll get bored downloading and building them by
hand in the right order. Here we recommend a much simpler way --- for the
full details of how to build Rivet by hand, please consult the Rivet Web page.

\paragraph{Bootstrap script}

We have written a bootstrapping
script which will download tarballs of Rivet, AGILe and the other required
libraries, expand them and build them in the right order with the correct build
flags. This is generally nicer than doing it all by hand, and virtually
essential if you want to use the existing versions of FastJet, HepMC, generator
libraries, and so on from CERN AFS: there are issues with these versions which
the script works around, which you won't find easy to do yourself.

To run the script, we recommend that you choose a personal installation
directory, i.e.\ make a \kbd{\home/local} directory for this purpose, to
avoid polluting your home directory with a lot of files. If you already use a
directory of the same name, you might want to use a separate one, say
\kbd{\home/rivetlocal}, such that if you need to delete everything in the
installation area you can do so without difficulties.

Now, change directory to your build area (you may also want to make this,
e.g. \kbd{\home/build}), and download the script:\\
\inp{wget \url{http://rivet.hepforge.org/svn/bootstrap/rivet-bootstrap}}\\
\inp{chmod +x rivet-bootstrap}\\
Now run it to get some help:
\inp{./rivet-bootstrap --help}\\
Now to actually do the install: for example, to bootstrap Rivet and AGILe
to the install area specified as the prefix argument, run this:\\
\inp{./rivet-bootstrap --install-agile --prefix=\val{localdir}}

If you are running on a system where the CERN AFS area is mounted as
\path{/afs/cern.ch}, then the bootstrap script will attempt to use the pre-built
HepMC\cite{Dobbs:2001ck}, LHAPDF\cite{Whalley:2005nh},
FastJet\cite{Cacciari:2005hq,fastjetweb} and GSL libraries from the LCG software area.
Either way, finally the bootstrap script will write out a file containing the
environment settings which will make the system useable. You can source this
file, e.g. \kbd{source rivetenv.sh} to make your current shell ready-to-go
for a Rivet run (use \kbd{rivetenv.csh} if you are a C shell user).

You now have a working, installed copy of the Rivet and AGILe libraries, and the
\kbd{rivet} and \kbd{agile-runmc} executables: respectively these are the
command-line frontend to the Rivet analysis library, and a convenient steering
command for generators which do not provide their own main program with HepMC
output. To test that they work as expected, source the setup scripts as above,
if you've not already done so, and run this:\\
\inp{rivet --help}\\
This should print a quick-reference user guide for the \kbd{rivet} command to
the terminal. Similarly, for \kbd{agile-runmc},\\
\inp{agile-runmc --help}\\
\inp{agile-runmc --list-gens}\\
\inp{agile-runmc --beams=pp:14000 Pythia6:425}\\
which should respectively print the help, list the available generators and make
10 LHC-type events using the Fortran Pythia\cite{Sjostrand:2006za} 6.423 generator. You're on your
way! If no generators are listed, you probaby need to install a local
Genser-type generator repository: see \SectionRef{sec:genser}.

In this manual, because of its convenience, we will use \kbd{agile-runmc} as our
canonical way of producing a stream of HepMC event data; if your interest is in
running a generator like Sherpa\cite{Gleisberg:2008ta},
Pythia~8\cite{Sjostrand:2007gs,Sjostrand:2008vc}, or Herwig++\cite{Bahr:2008pv}
which provides their own native way to make HepMC output, or a generator like
PHOJET which is not currently supported by AGILe, then substitute the
appropriate command in what follows.  We'll discuss using these commands in
detail in \SectionRef{sec:agile-runmc}.

\subsection{Getting generators for AGILe}
\label{sec:genser}

One last thing before continuing, though: the generators themselves. Again, if
you're running on a system with the CERN LCG AFS area mounted, then
\kbd{agile-runmc} will attempt to automatically use the generators packaged by the
LCG Genser team.

Otherwise, you'll have to build your own mirror of the LCG generators. This
process is evolving with time, and so, rather than provide information in this
manual which will be outdated by the time you read it, we simply refer you to
the relevant page on the Rivet wiki:
\url{http://rivet.hepforge.org/trac/wiki/GenserMirror}.

If you are interested in using a generator not currently supported by AGILe,
which does not output HepMC events in its native state, then please contact the
authors (via the Rivet developer contact email address) and hopefully we can
help.

\subsection{Command completion}

A final installation point worth considering is using the supplied bash-shell
programmable completion setup for the \kbd{rivet} and \kbd{agile-runmc}
commands. Despite being cosmetic and semi-trivial, programmable completion makes
using \kbd{rivet} positively pleasant, especially since you no longer need to
remember the somewhat cryptic analysis names\footnote{Standard Rivet analyses
  have names which, as well as the publication date and experiment name,
  incorporate the 8-digit Spires/Inspire ID code.}!

To use programmable completion, source the appropriate files from the install
location:\\
\inp{. \val{localdir}/share/Rivet/rivet-completion}\\
\inp{. \val{localdir}/share/AGILe/agile-completion}\\
(if you are using the setup script \kbd{rivetenv.sh} this is automatically
done for you).
If there is already a \kbd{\val{localdir}/etc/bash_completion.d} directory in
your install path, Rivet and AGILe's installation scripts will install extra
copies into that location, since automatically sourcing all completion files in
such a path is quite standard.


\section{Running Rivet analyses}
\label{sec:agile-runmc}

The \kbd{rivet} executable is the easiest way to use Rivet, and will be our
example throughout this manual. This command reads HepMC events in the standard
ASCII format, either from file or from a text stream.

\subsection{The FIFO idiom}
\label{sec:fifo-idiom}

Since you rarely want to store simulated HepMC events and they are
computationally cheap to produce (at least when compared to the remainder of
experiment simulation chains), we recommend using a Unix \emph{named pipe} (or
``FIFO'' --- first-in, first-out) to stream the events. While this may seem
unusual at first, it is just a nice way of ``pretending'' that we are writing to
and reading from a file, without actually involving any slow disk access or
building of huge files: a 1M event LHC run would occupy $\sim 60 GB$ on disk,
and typically it takes twice as long to make and analyse the events when the
filesystem is involved! Here is an example:\\
\inp{mkfifo fifo.hepmc}\\
\inp{agile-runmc Pythia6:425 -o fifo.hepmc \&}\\
\inp{rivet -a EXAMPLE fifo.hepmc}\\
Note that the generator process (\kbd{agile-runmc} in this case) is
\emph{backgrounded} before \kbd{rivet} is run.

Notably, \kbd{mkfifo} will not work if applied to a directory mounted via the
AFS distributed filesystem, as widely used in HEP. This is not a big problem:
just make your FIFO object somewhere not mounted via AFS, e.g. \kbd{/tmp}. There
is no performance penalty, as the filesystem object is not written to during the
streaming process.

In the following command examples, we will assume that a generator has been set
up to write to the \kbd{fifo.hepmc} FIFO, and just list the \kbd{rivet} command
that reads from that location. Some typical \kbd{agile-runmc} commands are
listed in \AppendixRef{app:agilerunmc}.

\subsection{Analysis status}

The standard Rivet analyses are divided into four status classes: validated,
preliminary, obsolete, and unvalidated (in roughly decreasing order of academic
acceptability).

The Rivet ``validation procedure'' is not formally
defined, but generally implies that an analysis has been checked to ensure
reproduction of MC points shown in the paper where possible, and is believed to
have no outstanding issues with analysis procedure or cuts.  Additionally,
analyses marked as ``validated'' and distributed with Rivet should normally have
been code-checked by an experienced developer to ensure that the code is a good
example of Rivet usage and is not more complex than required or otherwise
difficult to read or maintain. Such analyses are regarded as fully ready for use
in any MC validation or tuning studies.

Validated analyses which implement an unfinished piece of experimental work are
considered to be trustworthy in their implementation of a conference note or
similar ``informal'' publication, but do not have the magic stamp of approval
that comes from a journal publication. This remains the standard mark of
experimental respectability and accordingly we do not include such analyses in
the Rivet standard analysis libraries, but in a special ``preliminary''
library. While preliminary analyses may be used for physics studies, please be
aware of the incomplete status of the corresponding experimental study, and also
be aware that the histograms in such analyses may be renamed or removed
entirely, as may the analysis itself.

Preliminary analyses will not have a Spires/Inspire ID, and hence on their
move into the standard Rivet analysis library they will normally undergo a name
change: please ensure when you upgrade between Rivet versions that any scripts
or programs which were using preliminary analyses are not broken by the
disappearance or change of that analysis in the newer version. The minor perils
of using preliminary analyses can be avoided by the cautious by building Rivet
with the \kbd{-{}-disable-preliminary} configuration flag, in which case their
temptation will not even be offered.

To make transitions between Rivet versions more smooth and predictable for users
of preliminary analyses, preliminary analyses which are superseded by a
validated version will be reclassified as obsolete and will be retained for one
major version of Rivet with a status of "obsolete" before being removed, to give
users time to migrate their run scripts, i.e. if an analysis is marked as
obsolete in version 1.4.2, it will remain in Rivet's distribution until version
1.5.0.  Obsolete analyses may have different reference histograms from the final
version and will not be maintained. Obsolete analyses will not be built if
either the \kbd{-{}-disable-obsolete} configuration flag is specified at build
time: for convenience, the default value of this flag is the value of the
\kbd{-{}-disable-preliminary} flag.

Finally, unvalidated analyses are those whose implementation is incomplete,
flawed or just troubled by doubts. Running such analyses is not a good idea if
you aren't trying to fix them, and Rivet's command line tools will print copious
warning messages if you do. Unvalidated analyses in the Rivet distribution are
not built by default, as they are only of interest to developers and would be
distracting clutter for the majority of users: if you \emph{really} need them,
building Rivet with the \kbd{-{}-enable-unvalidated} configuration flag will
slake your thirst for danger.

\subsection{Example \kbd{rivet} commands}

\begin{itemize}

\item \textbf{Getting help: }{\kbd{rivet --help} will print a (hopefully)
    helpful list of options which may be used with the \kbd{rivet} command, as
    well as other information such as environment variables which may affect the
    run.}

\item \textbf{Choosing analyses: }{\kbd{rivet --list-analyses} will list the
    available analyses, including both those in the Rivet distribution and any
    plugins which are found at runtime. \kbd{rivet --show-analysis \val{patt}}
    will show a lot of details about any analyses whose name match the
    \val{patt} regular expression pattern --- simple bits of analysis name are a
    perfectly valid subset of this. For example, \kbd{rivet --show-analysis
      CDF_200} exploits the standard Rivet analysis naming scheme to show
    details of all available CDF experiment analyses published in the
    ``noughties.''}

\item \textbf{Running analyses: }{\kbd{rivet -a~DELPHI_1996_S3430090
      fifo.hepmc} will run the Rivet
    \kbd{DELPHI_1996_S3430090}\cite{Abreu:1996na} analysis on the events in the
    \kbd{fifo.hepmc} file (which, from the name, is probably a filesystem named
    pipe rather than a normal \emph{file}). This analysis is the one originally
    used for the \Delphi ``\textsc{Professor}'' generator tuning. If the first
    event in the data file does not have appropriate beam particles, the
    analysis will be disabled; since there is only one analysis in this case,
    the command will exit immediately with a warning if the first event is not
    an $\Ppositron\Pelectron$ event.}


\item \textbf{Histogramming: }{\kbd{rivet fifo.hepmc -H~foo.aida} will read all the
    events in the \kbd{fifo.hepmc} file. The \kbd{-H} switch is used to
    specify that the output histogram file will be named \kbd{foo.aida}. By
    default the output file is called \kbd{Rivet.aida}.}

\item \textbf{Fine-grained logging: }

  {\kbd{rivet fifo.hepmc -A
      -l~Rivet.Analysis=DEBUG~\cmdbreak -l~Rivet.Projection=DEBUG
      -l~Rivet.Projection.FinalState=TRACE~\cmdbreak -l~NEvt=WARN}
    will analyse events as before, but will print different status
    information as the run progresses. Hierarchical logging control is possible
    down to the level of individual analyses and projections as shown above;
    this is useful for debugging without getting overloaded with debug
    information from \emph{all} the components at once. The default level is
    ``\textsc{info}'', which lies between ``\textsc{debug}'' and
    ``\textsc{warning}''; the ``\textsc{trace}'' level is for very low level
    information, and probably isn't needed by normal users.}

\end{itemize}

\section{Using analysis data}

In this section, we summarise how to use the data files which Rivet produces for
plotting, validation and tuning.

\subsection{Histogram formats}

Rivet currently produces output histogram data in the AIDA XML format. Most
people aren't familiar with AIDA (and we recommend that you remain that way!),
and it will disappear entirely from Rivet in version 2.0. If you do not want to
use the plotting tools that come with Rivet (cf.\ Sec.~\ref{sec:plotting}), you might
wish to cast the AIDA files to a different format for plotting, and for this we
supply several scripts.

\paragraph{Conversion to ROOT}

Rivet installs an
\kbd{aida2root} script, which converts the AIDA records to a \kbd{.root} file
full of ROOT \texttt{TGraph}s. One word of warning: a bug in ROOT means that
\texttt{TGraph}s do not render properly from file because the axis is not drawn by
default. To display the plots correctly in ROOT you will need to pass the
\kbd{"AP"} drawing option string to either the \kbd{TGraph::Draw()} method, or
in the options box in the \kbd{TBrowser} GUI interface.
Alternatively you can also use the \kbd{-t} option with which \kbd{aida2root}
produces \texttt{TH1}s instead.

\paragraph{Conversion to ``flat format''}

Most of our histogramming is based around a ``flat'' plain text format,
which can easily be read (and written) by hand. We provide a script called
\kbd{aida2flat} to do this conversion. Run \kbd{aida2flat -h} to get usage
instructions; in particular the Gnuplot and ``split output'' options are useful
for further visualisation. Aside from anything else, this is useful for simply
checking the contents of an AIDA file, with \kbd{aida2flat Rivet.aida | less}.

\vspace{1.8em}

\begin{detail}
  We get asked a lot about why we don't use ROOT internally: aside from a
  general unhappiness about the design and quality of the data objects in ROOT,
  the monolithic nature of the system makes it a big dependency for a system as
  small as Rivet. While not an issue for experimentalists, most theorists and
  generator developers do not use ROOT and we preferred to embed the AIDA
  system, which in its LWH implementation requires no external package. The
  replacement for AIDA will be another lightweight system rather than ROOT, with
  an emphasis on friendly, intuitive data object design, and correct handling of
  sample merging statistics for all data objects.
\end{detail}

\subsection{Chopping histograms}
\newcommand{\chophisto}{\kbd{rivet-chopbins}\xspace} In some cases you don't
want to keep the complete histograms produced by Rivet.  For generator tuning
purposes, for example, you want to get rid of the bins you already know your
generator is incapable of describing. You can use the script \chophisto to
specify those bin-ranges you want to keep individually for each histogram in a
Rivet output-file. The bin-ranges have to be specified using the corresponding
x-values of that histogram.  The usage is very simple. You can specify bin
ranges of histograms to keep on the command-line via the \kbd{-b}
switch, which can be given multiple times, e.g.\\
\kbd{\chophisto -b /CDF\_2001\_S4751469/d03-x01-y01:5:13 Rivet.aida}\\
will chop all bins with $x<5$ and $x>13$ from the histogram
\kbd{/CDF\_2001\_S4751469/d03\-x01\-y01} in the file \kbd{Rivet.aida}. (In
this particular case, $x$ would be a leading jet \pT.)

\subsection{Normalising histograms}
\newcommand{\normhisto}{\kbd{rivet-rescale }} Sometimes you want to
use histograms normalised to, e.g., the generator cross-section or the area of
a reference-data histogram. The script \normhisto was designed for these
purposes. The usage is the following:\\
\kbd{\normhisto -O observables -r RIVETDATA -o normalised Rivet.aida}\\
By default, the normalised histograms are written to file in the AIDA-XML
format. You can also give the \kbd{-f} switch on the command line to produce
flat histograms.

\paragraph{Normalising to reference data} You will need an output-file of
Rivet, \kbd{Rivet.aida}, a folder that contains the reference-data histograms
(e.g. \kbd{rivet-config --datadir}) and optionally, a text-file,
\kbd{observables} that contains the names of the histograms you would like to
normalise - those not given in the file will remain un-normalised. These
are examples of how your \kbd{observables} file might look like:
\begin{snippet}
/CDF_2000_S4155203/d01-x01-y01
\end{snippet}

If a histogram \kbd{/CDF\_2000\_S4155203/d01-x01-y01} is found in one of the
reference-data files in the folder specified via the \kbd{-r} switch, then this
will result in a histogram \kbd{/CDF\_2000\_S4155203/d01-x01-y01} being
normalised to the area of the corresponding reference-data histogram.  You can
further specify a certain range of bins to normalise:
\begin{snippet}
/CDF_2000_S4155203/d01-x01-y01:2:35
\end{snippet}
\noindent will chop off the bins
with $x<2$ and $x>35$ of both, the histogram in your \kbd{Rivet.aida} and the
reference-data histogram. The remaining MC histogram is then normalised to the
remaining area of the reference-data histogram.

\paragraph{Normalising to arbitrary areas}%
In the file \kbd{observables} you
can further specify an arbitrary number, e.g. a generator cross-section, as
follows:
\begin{snippet}
/CDF_2000_S4155203/d01-x01-y01  1.0
\end{snippet}
\noindent will result in the histogram \kbd{/CDF\_2000\_S4155203/d01-x01-y01} being
normalised to 1.0, and
\begin{snippet}
/CDF_2000_S4155203/d01-x01-y01:2:35  1.0
\end{snippet}
\noindent will chop off the bins with $x<2$ and $x>35$ of the histogram\\
\kbd{/CDF\_2000\_S4155203/d01-x01-y01} first and normalise the remaining
histogram to one.

\subsection{Plotting and comparing data}
\label{sec:plotting}
Rivet comes with three commands --- \kbd{rivet-mkhtml}, \kbd{compare-histos} and
\kbd{make-plots} --- for comparing and plotting data files. These commands
produce nice comparison plots of publication quality from the AIDA format text
files.

The high level program \kbd{rivet-mkhtml} will automatically create a plot
webpage from the given AIDA files. It searches for reference data automatically
and uses the other two commands internally. Example:\\
\inp{rivet-mkhtml withUE.aida:'Title=With UE' withoutUE.aida:'LineColor=blue'}\\
Run \kbd{rivet-mkhtml --help} to find out about all features and options.

You can also run the other two commands separately:
\begin{itemize}
\item \kbd{compare-histos} will accept a number of AIDA files as input (ending in
\kbd{.aida}), identify which plots are available in them, and combine the MC
and reference plots appropriately into a set of plot data files ending with
\kbd{.dat}. More options are described by running \kbd{compare-histos --help}.

Incidentally, the reference files for each Rivet analysis are to be found in the
installed Rivet shared data directory, \kbd{\val{installdir}/share/Rivet}. You
can find the location of this by using the \kbd{rivet-config} command:\\
\inp{rivet-config --datadir}

\item You can plot the created data files using the \kbd{make-plots} command:\\
\inp{make-plots --pdf *.dat}\\
The \kbd{--pdf} flag makes the output plots in PDF format: by default the output
is in PostScript (\kbd{.ps}), and flags for conversion to EPS and PNG are also
available.
\end{itemize}

\cleardoublepage

\part{Selected analyses}
\label{part:selectedanalyses}
\makeatletter
\renewcommand{\d}[1]{\ensuremath{\mathrm{#1}}}
\let\old@eta\eta
\renewcommand{\eta}{\ensuremath{\old@eta}\xspace}
\let\old@phi\phi
\renewcommand{\phi}{\ensuremath{\old@phi}\xspace}
\providecommand{\pT}{\ensuremath{p_\perp}\xspace}
\providecommand{\pTmin}{\ensuremath{p_\perp^\text{min}}\xspace}
\makeatother

Each Rivet release is accompanied by a standard library of analyses
implementing currently a total of 250 experimental measurements or Monte-Carlo validation
studies. The full listing of these is beyond the scope of this publication, but
it is available both online at \url{http://rivet.hepforge.org/analyses} and as
a part of the manual coming with each release of Rivet in the \kbd{doc/}
sub-directory. Here, we only want to show-case a selection of analyses spanning the full
spectrum of experiments from LEP over HERA to Tevatron and the LHC and
demonstrating the versatility of the Rivet framework.

For each of the 250 analyses, in addition to a brief summary one can find
information about
the collider at which the measurement was made, references to the original
publications, status and authors of the Rivet implementation as well as run
details necessary for comparing a Monte-Carlo prediction with the data.

\section{Selection of analyses available in the Rivet framework}
\subsection[ALEPH\_1996\_S3196992]{ALEPH\_1996\_S3196992\,\cite{Buskulic:1995au}:\\ Measurement of the quark to photon fragmentation function}

\noindent Earlier measurements at LEP of isolated hard photons in hadronic Z decays, attributed to radiation from primary quark pairs, have been extended in the ALEPH experiment to include hard photon production inside hadron jets. Events are selected where all particles combine democratically to form hadron jets, one of which contains a photon with a fractional energy $z > 0.7$. After statistical subtraction of non-prompt photons, the quark-to-photon fragmentation function, $D(z)$, is extracted directly from the measured 2-jet rate.

\noindent\textsc{Beams:} $e^+$\,$e^-$ \newline
\textsc{Energies:} (45.6, 45.6) GeV \newline
\textsc{Experiment:} ALEPH (LEP Run 1)\newline
\textsc{Spires ID:} \href{http://inspire-hep.net/search?p=find+key+3196992}{3196992}\newline
\textsc{Status:} VALIDATED\newline
\textsc{Authors:}
 \penalty 100
\begin{itemize}
  \item Frank Siegert $\langle\,$\href{mailto:frank.siegert@cern.ch}{frank.siegert@cern.ch}$\,\rangle$
\end{itemize}
\textsc{Run details:}
 \penalty 100
\begin{itemize}

  \item $e^+e^-\to$ jets with $\pi$ and $\eta$ decays turned off.\end{itemize}

\subsection[ALICE\_2011\_S8945144]{ALICE\_2011\_S8945144\,\cite{Aamodt:2011my}:\\ Tranverse momentum spectra of pions, kaons and protons in pp collisions at 0.9 TeV}

\noindent Obtaining the tranverse momentum spectra of pions, kaons and protons in $pp$ collisions at $\sqrt{s} = 0.9$ TeV with ALICE at the LHC. Mean transverse momentum as a function of the mass of the emitted particle is also included.

\noindent\textsc{Beams:} $p$\,$p$ \newline
\textsc{Energies:} (450.0, 450.0) GeV \newline
\textsc{Experiment:} ALICE (LHC)\newline
\textsc{Spires ID:} \href{http://inspire-hep.net/search?p=find+key+8945144}{8945144}\newline
\textsc{Status:} VALIDATED\newline
\textsc{Authors:}
 \penalty 100
\begin{itemize}
  \item Pablo Bueno Gomez $\langle\,$\href{mailto:UO189399@uniovi.es}{UO189399@uniovi.es}$\,\rangle$
  \item Eva Sicking $\langle\,$\href{mailto:esicking@cern.ch}{esicking@cern.ch}$\,\rangle$
\end{itemize}
\textsc{Run details:}
 \penalty 100
\begin{itemize}

  \item Diffractive events need to be enabled.\end{itemize}

\subsection[ARGUS\_1993\_S2653028]{ARGUS\_1993\_S2653028\,\cite{Albrecht:1992qf}:\\ Inclusive production of charged pions, kaons and protons in $\Upsilon(4S)$ decays.}

\noindent Measurement of inclusive production of charged pions, kaons and protons from $\Upsilon(4S)$ decays. Kaon spectra are determined in two different ways using particle identification and detecting decays in-flight. Results are background continuum subtracted. This analysis is useful for tuning $B$ meson decay modes.

\noindent\textsc{Beams:} $e^+$\,$e^-$ \newline
\textsc{Energies:} (5.3, 5.3) GeV \newline
\textsc{Spires ID:} \href{http://inspire-hep.net/search?p=find+key+2653028}{2653028}\newline
\textsc{Status:} VALIDATED\newline
\textsc{Authors:}
 \penalty 100
\begin{itemize}
  \item Peter Richardson $\langle\,$\href{mailto:Peter.Richardson@durham.ac.uk}{Peter.Richardson@durham.ac.uk}$\,\rangle$
\end{itemize}
\textsc{Run details:}
 \penalty 100
\begin{itemize}

  \item $e^+ e^-$ analysis on the $\Upsilon(4S)$ resonance.\end{itemize}

\subsection[ATLAS\_2012\_I1094568]{ATLAS\_2012\_I1094568\,\cite{ATLAS:2012al}:\\ Measurement of ttbar production with a veto on additional central jet activity}

\noindent A measurement of the additional jet activity in dileptonic ttbar events. The fraction of events passing a veto requirement are shown as a function the veto scale for four central rapidity intervals. Two veto definitions are used: events are vetoed if they contain an additional jet in the rapidity interval with transverse momentum above a threshold, or alternatively, if the scalar transverse momentum sum of all additional jets in the rapidity interval is above a threshold.

\noindent\textsc{Beams:} $p$\,$p$ \newline
\textsc{Energies:} (3500.0, 3500.0) GeV \newline
\textsc{Experiment:} ATLAS (LHC)\newline
\textsc{Spires ID:} \href{http://inspire-hep.net/search?p=find+key+1094568}{1094568}\newline
\textsc{Status:} VALIDATED\newline
\textsc{Authors:}
 \penalty 100
\begin{itemize}
  \item Kiran Joshi $\langle\,$\href{mailto:kiran.joshi@cern.ch}{kiran.joshi@cern.ch}$\,\rangle$
\end{itemize}
\textsc{Run details:}
 \penalty 100
\begin{itemize}

  \item Require dileptonic ttbar events at 7TeV.\end{itemize}

\subsection[BABAR\_2007\_S7266081]{BABAR\_2007\_S7266081\,\cite{Aubert:2007mh}:\\ Measurements of Semi-Leptonic Tau Decays into Three Charged Hadrons}

\noindent Measurement of tau decays to three charged hadrons using a data sample corresponding to an integrated luminosity of 342 $fb^{-1}$ collected with the BABAR detector at the SLAC PEP-II electron-positron storage ring operating at a center-of-mass energy near 10.58 GeV.

\noindent\textsc{Beams:} $e^+$\,$e^-$ \newline
\textsc{Energies:} (3.5, 8.0) GeV \newline
\textsc{Spires ID:} \href{http://inspire-hep.net/search?p=find+key+7266081}{7266081}\newline
\textsc{Status:} VALIDATED\newline
\textsc{Authors:}
 \penalty 100
\begin{itemize}
  \item Peter Richardson $\langle\,$\href{mailto:Peter.Richardson@durham.ac.uk}{Peter.Richardson@durham.ac.uk}$\,\rangle$
\end{itemize}
\textsc{Run details:}
 \penalty 100
\begin{itemize}

  \item Tau production, can be any process but original data was in  $e^+ e^-$ at the $\Upsilon(4S)$ resonance, with CoM boost -- 8.0~GeV~($e^−$) and 3.5~GeV~($e^+$)\end{itemize}

\subsection[BELLE\_2006\_S6265367]{BELLE\_2006\_S6265367\,\cite{Seuster:2005tr}:\\ Charm hadrons from fragmentation and B decays on the $\Upsilon(4S)$}

\noindent Analysis of charm quark fragmentation at 10.6 GeV, based on a data sample of 103 fb collected by the Belle detector at the KEKB accelerator. Fragmentation into charm is studied for the main charmed hadron ground states, namely $D^0$, $D^+$, $D^+_s$ and $\Lambda_c^+$, as well as the excited states $D^{*0}$ and $D^{*+}$. This analysis can be used to constrain charm fragmentation in Monte Carlo generators. As the original data are not corrected for the branching ratios of the decay modes used to observed the charm hadrons we also include distributions with unit normalisation which are more useful for Monte Carlo tuning.

\noindent\textsc{Beams:} $e^+$\,$e^-$ \newline
\textsc{Energies:} (3.5, 8.0), (3.5, 7.9) GeV \newline
\textsc{Spires ID:} \href{http://inspire-hep.net/search?p=find+key+6265367}{6265367}\newline
\textsc{Status:} VALIDATED\newline
\textsc{Authors:}
 \penalty 100
\begin{itemize}
  \item Jan Eike von Seggern $\langle\,$\href{mailto:jan.eike.von.seggern@physik.hu-berlin.de}{jan.eike.von.seggern@physik.hu-berlin.de}$\,\rangle$
\end{itemize}
\textsc{Run details:}
 \penalty 100
\begin{itemize}

  \item $e^+ e^-$ analysis on the $\Upsilon(4S)$ resonance, with CoM boost -- 8.0~GeV~($e^−$) and 3.5~GeV~($e^+$)\end{itemize}

\subsection[CDF\_2001\_S4751469]{CDF\_2001\_S4751469\,\cite{Affolder:2001xt}:\\ Field \& Stuart Run I underlying event analysis.}

\noindent The original CDF underlying event analysis, based on decomposing each event into a transverse structure with ``toward'', ``away'' and ``transverse'' regions defined relative to the azimuthal direction of the leading jet in the event. Since the toward region is by definition dominated by the hard process, as is the away region by momentum balance in the matrix element, the transverse region is most sensitive to multi-parton interactions. The transverse regions occupy $|\phi| \in [60\degree, 120\degree]$ for $|\eta| < 1$. The \pT ranges for the leading jet are divided experimentally into the `min-bias' sample from 0--20 GeV, and the `JET20' sample from 18--49 GeV.

\noindent\textsc{Beams:} $\bar{p}$\,$p$ \newline
\textsc{Energies:} (900.0, 900.0) GeV \newline
\textsc{Experiment:} CDF (Tevatron Run 1)\newline
\textsc{Spires ID:} \href{http://inspire-hep.net/search?p=find+key+4751469}{4751469}\newline
\textsc{Status:} VALIDATED\newline
\textsc{Authors:}
 \penalty 100
\begin{itemize}
  \item Andy Buckley $\langle\,$\href{mailto:andy.buckley@cern.ch}{andy.buckley@cern.ch}$\,\rangle$
\end{itemize}
\textsc{Run details:}
 \penalty 100
\begin{itemize}

  \item $p\bar{p}$ QCD interactions at 1800 GeV. The leading jet is binned from 0--49 GeV, and histos can usually can be filled with a single generator run without kinematic sub-samples.\end{itemize}

\subsection[CLEO\_2004\_S5809304]{CLEO\_2004\_S5809304\,\cite{Artuso:2004pj}:\\ Charm hadrons from fragmentation near the $\Upsilon(4S)$}

\noindent Analysis of charm quark fragmentation at 10.5 GeV, based on a data sample of 103 fb collected by the CLEO experiment. Fragmentation into charm is studied for the charmed hadron ground states, namely $D^0$, $D^+$, as well as the excited states $D^{*0}$ and $D^{*+}$. This analysis can be used to constrain charm fragmentation in Monte Carlo generators.

\noindent\textsc{Beams:} $e^+$\,$e^-$ \newline
\textsc{Energies:} (5.3, 5.3) GeV \newline
\textsc{Spires ID:} \href{http://inspire-hep.net/search?p=find+key+6265367}{6265367}\newline
\textsc{Status:} VALIDATED\newline
\textsc{Authors:}
 \penalty 100
\begin{itemize}
  \item Peter Richardson $\langle\,$\href{mailto:Peter.Richardson@durham.ac.uk}{Peter.Richardson@durham.ac.uk}$\,\rangle$
\end{itemize}
\textsc{Run details:}
 \penalty 100
\begin{itemize}

  \item $e^+ e^-$ analysis near the $\Upsilon(4S)$ resonance\end{itemize}

\subsection[CMS\_2011\_S8957746]{CMS\_2011\_S8957746\,\cite{Khachatryan:2011dx}:\\ Event shapes}

\noindent Central transverse Thrust and Minor have been measured in proton-proton collisions at \ensuremath{\sqrt{s}}=7 TeV, with a data sample collected with the CMS detector at the LHC. The sample corresponds to an integrated luminosity of 3.2 inverse picobarns. Input for the variables are anti-kt jets with $R=0.5$.

\noindent\textsc{Beams:} $p$\,$p$ \newline
\textsc{Energies:} (3500.0, 3500.0) GeV \newline
\textsc{Experiment:} CMS (LHC)\newline
\textsc{Spires ID:} \href{http://inspire-hep.net/search?p=find+key+8957746}{8957746}\newline
\textsc{Status:} VALIDATED\newline
\textsc{Authors:}
 \penalty 100
\begin{itemize}
  \item Hendrik Hoeth $\langle\,$\href{mailto:hendrik.hoeth@cern.ch}{hendrik.hoeth@cern.ch}$\,\rangle$
\end{itemize}
\textsc{Run details:}
 \penalty 100
\begin{itemize}

  \item pp QCD interactions at 7000 GeV. Particles with c*tau>10mm are stable.\end{itemize}

\subsection[D0\_2008\_S7719523]{D0\_2008\_S7719523\,\cite{Abazov:2008er}:\\ Isolated $\gamma$ + jet cross-sections, differential in \pT($\gamma$) for various $y$ bins}

\noindent The process $p \bar{p}$ \ensuremath{\to} photon + jet + X as studied by the D0 detector at the Fermilab Tevatron collider at center-of-mass energy \ensuremath{\sqrt{s}} = 1.96 TeV. Photons are reconstructed in the central rapidity region $|y_\gamma| < 1.0$ with transverse momenta in the range 30--400 GeV, while jets are reconstructed in either the central $|y_\text{jet}| < 0.8$ or forward $1.5 < |y_\text{jet}| < 2.5$ rapidity intervals with $\pT^\text{jet} > 15~\text{GeV}$. The differential cross section $\mathrm{d}^3 \sigma / \mathrm{d}{\pT^\gamma} \mathrm{d}{y_\gamma} \mathrm{d}{y_\text{jet}}$ is measured as a function of $\pT^\gamma$ in four regions, differing by the relative orientations of the photon and the jet.  MC predictions have trouble with simultaneously describing the measured normalization and $\pT^\gamma$ dependence of the cross section in any of the four measured regions.

\noindent\textsc{Beams:} $\bar{p}$\,$p$ \newline
\textsc{Energies:} (980.0, 980.0) GeV \newline
\textsc{Experiment:} D0 (Tevatron Run 2)\newline
\textsc{Spires ID:} \href{http://inspire-hep.net/search?p=find+key+7719523}{7719523}\newline
\textsc{Status:} VALIDATED\newline
\textsc{Authors:}
 \penalty 100
\begin{itemize}
  \item Andy Buckley $\langle\,$\href{mailto:andy.buckley@cern.ch}{andy.buckley@cern.ch}$\,\rangle$
  \item Gavin Hesketh $\langle\,$\href{mailto:gavin.hesketh@cern.ch}{gavin.hesketh@cern.ch}$\,\rangle$
  \item Frank Siegert $\langle\,$\href{mailto:frank.siegert@cern.ch}{frank.siegert@cern.ch}$\,\rangle$
\end{itemize}
\textsc{Run details:}
 \penalty 100
\begin{itemize}

  \item Produce only gamma + jet ($q,\bar{q},g$) hard processes (for Pythia 6, this means MSEL=10 and MSUB indices 14, 29 \& 115 enabled). The lowest bin edge is at 30 GeV, so a kinematic \pTmin cut is probably required to fill the histograms.\end{itemize}

\subsection[DELPHI\_1996\_S3430090]{DELPHI\_1996\_S3430090\,\cite{Abreu:1996na}:\\ Delphi MC tuning on event shapes and identified particles.}

\noindent Event shape and charged particle inclusive distributions measured using 750000 decays of Z bosons to hadrons from the DELPHI detector at LEP. This data, combined with identified particle distributions from all LEP experiments, was used for tuning of shower-hadronisation event generators by the original PROFESSOR method.  This is a critical analysis for MC event generator tuning of final state radiation and both flavour and kinematic aspects of hadronisation models.

\noindent\textsc{Beams:} $e^+$\,$e^-$ \newline
\textsc{Energies:} (45.6, 45.6) GeV \newline
\textsc{Experiment:} DELPHI (LEP 1)\newline
\textsc{Spires ID:} \href{http://inspire-hep.net/search?p=find+key+3430090}{3430090}\newline
\textsc{Status:} VALIDATED\newline
\textsc{Authors:}
 \penalty 100
\begin{itemize}
  \item Andy Buckley $\langle\,$\href{mailto:andy.buckley@cern.ch}{andy.buckley@cern.ch}$\,\rangle$
  \item Hendrik Hoeth $\langle\,$\href{mailto:hendrik.hoeth@cern.ch}{hendrik.hoeth@cern.ch}$\,\rangle$
\end{itemize}
\textsc{Run details:}
 \penalty 100
\begin{itemize}

  \item \ensuremath{\sqrt{s}} = 91.2 GeV, $e^+ e^- \ensuremath{\to} Z^0$ production with hadronic decays only\end{itemize}

\subsection[H1\_2000\_S4129130]{H1\_2000\_S4129130\,\cite{Adloff:1999ws}:\\ H1 energy flow in DIS}

\noindent Measurements of transverse energy flow for neutral current deep- inelastic scattering events produced in positron-proton collisions at HERA. The kinematic range covers squared momentum transfers $Q^2$ from 3.2 to 2200 GeV$^2$; the Bjorken scaling variable $x$ from $8 \times 10^{-5}$ to 0.11 and the hadronic mass $W$ from 66 to 233 GeV. The transverse energy flow is measured in the hadronic centre of mass frame and is studied as a function of $Q^2$, $x$, $W$ and pseudorapidity. The behaviour of the mean transverse energy in the central pseudorapidity region and an interval corresponding to the photon fragmentation region are analysed as a function of $Q^2$ and $W$.  This analysis is useful for exploring the effect of photon PDFs and for tuning models of parton evolution and treatment of fragmentation and the proton remnant in DIS.

\noindent\textsc{Beams:} $p$\,$e^+$ \newline
\textsc{Energies:} (820.0, 27.5) GeV \newline
\textsc{Experiment:} H1 (HERA)\newline
\textsc{Spires ID:} \href{http://inspire-hep.net/search?p=find+key+4129130}{4129130}\newline
\textsc{Status:} VALIDATED\newline
\textsc{Authors:}
 \penalty 100
\begin{itemize}
  \item Peter Richardson $\langle\,$\href{mailto:peter.richardson@durham.ac.uk}{peter.richardson@durham.ac.uk}$\,\rangle$
\end{itemize}
\textsc{Run details:}
 \penalty 100
\begin{itemize}

  \item $e^+ p$ deep inelastic scattering with $p$ at 820 GeV, $e^+$ at 27.5 GeV \ensuremath{\to} \ensuremath{\sqrt{s}} = 300 GeV\end{itemize}

\subsection[JADE\_1998\_S3612880]{JADE\_1998\_S3612880\,\cite{MovillaFernandez:1997fr}:\\ Event shapes for 22, 35 and 44 GeV}

\noindent Thrust, Jet Mass and Broadenings, Y23 for 35 and 44 GeV and only Y23 at 22 GeV.

\noindent\textsc{Beams:} $e^-$\,$e^+$ \newline
\textsc{Energies:} (11.0, 11.0), (17.5, 17.5), (22.0, 22.0) GeV \newline
\textsc{Experiment:} JADE (PETRA)\newline
\textsc{Spires ID:} \href{http://inspire-hep.net/search?p=find+key+3612880}{3612880}\newline
\textsc{Status:} VALIDATED\newline
\textsc{Authors:}
 \penalty 100
\begin{itemize}
  \item Holger Schulz $\langle\,$\href{mailto:holger.schulz@physik.hu-berlin.de}{holger.schulz@physik.hu-berlin.de}$\,\rangle$
\end{itemize}
\textsc{Run details:}
 \penalty 100
\begin{itemize}

  \item Z\ensuremath{\to}hadronic final states, bbar contributions have been corrected for as well as ISR\end{itemize}

\subsection[LHCB\_2011\_I919315]{LHCB\_2011\_I919315\,\cite{Aaij:2011uk}:\\ Inclusive differential $\Phi$ production cross-section as a function of $p_\text{T}$ and $y$}

\noindent Measurement of the inclusive differential $\Phi$ cross-section in $pp$ collisions at $\sqrt {s}=7$TeV in the rapidity range of $ 2.44 < y < 4.06$ and the $p_\text{T}$ range of 0.6 GeV/c $< p_\text{T} <$ 5.0 GeV/c.

\noindent\textsc{Beams:} $p$\,$p$ \newline
\textsc{Energies:} (3500.0, 3500.0) GeV \newline
\textsc{Experiment:} LHCB (LHC)\newline
\textsc{Status:} VALIDATED\newline
\textsc{Authors:}
 \penalty 100
\begin{itemize}
  \item Friederike Blatt $\langle\,$\href{mailto:friederike.blatt@tu-dortmund.de}{friederike.blatt@tu-dortmund.de}$\,\rangle$
  \item Michael Kaballo $\langle\,$\href{mailto:michael.kaballo@tu-dortmund.de}{michael.kaballo@tu-dortmund.de}$\,\rangle$
  \item Till Moritz Karbach $\langle\,$\href{mailto:moritz.karbach@tu-dortmund.de}{moritz.karbach@tu-dortmund.de}$\,\rangle$
\end{itemize}
\textsc{Run details:}
 \penalty 100
\begin{itemize}

  \item $pp$ collisions, QCD-Events, $\sqrt{s}=7$TeV\end{itemize}

\subsection[LHCF\_2012\_I1115479]{LHCF\_2012\_I1115479\,\cite{Adriani:2012ap}:\\ Measurement of forward neutral pion transverse momentum spectra for $\sqrt{s}$ = 7 TeV proton-proton collisions at LHC}

\noindent The inclusive production rate of neutral pions has been measured by LHCf experiment during $\sqrt{s}=7$ TeV pp collision operation in early 2010. In order to ensure good event reconstruction efficiency, the range of the $\pi^0$ rapidity and $p_\perp$ are limited to $8.9 < y < 11.0$ and $p_\perp < 0.6$ GeV, respectively.

\noindent\textsc{Beams:} $p$\,$p$ \newline
\textsc{Energies:} (3500.0, 3500.0) GeV \newline
\textsc{Experiment:} LHCF (LHC)\newline
\textsc{Status:} VALIDATED\newline
\textsc{Authors:}
 \penalty 100
\begin{itemize}
  \item Sercan Sen $\langle\,$\href{mailto:ssen@cern.ch}{ssen@cern.ch}$\,\rangle$
\end{itemize}
\textsc{Run details:}
 \penalty 100
\begin{itemize}

  \item Inelastic events (ND+SD+DD) at $\sqrt{s}$ = 7 TeV.\end{itemize}

\subsection[OPAL\_2004\_S6132243]{OPAL\_2004\_S6132243\,\cite{Abbiendi:2004qz}:\\ Event shape distributions and moments in $e^+ e^-$ \ensuremath{\to} hadrons at 91--209 GeV}

\noindent Measurement of $e^+ e^-$ event shape variable distributions and their 1st to 5th moments in LEP running from the Z pole to the highest LEP 2 energy of 209 GeV.

\noindent\textsc{Beams:} $e^+$\,$e^-$ \newline
\textsc{Energies:} (45.6, 45.6), (66.5, 66.5), (88.5, 88.5), (98.5, 98.5) GeV \newline
\textsc{Experiment:} OPAL (LEP 1 \& 2)\newline
\textsc{Spires ID:} \href{http://inspire-hep.net/search?p=find+key+6132243}{6132243}\newline
\textsc{Status:} VALIDATED\newline
\textsc{Authors:}
 \penalty 100
\begin{itemize}
  \item Andy Buckley $\langle\,$\href{mailto:andy.buckley@cern.ch}{andy.buckley@cern.ch}$\,\rangle$
\end{itemize}
\textsc{Run details:}
 \penalty 100
\begin{itemize}

  \item Hadronic $e^+ e^-$ events at 4 representative energies (91, 133, 177, 197). Runs need to have ISR suppressed, since the analysis was done using a cut of $\sqrt{s} - \sqrt{s_\text{reco}} < 1\,\text{GeV}$. Particles with a livetime $> 3 \cdot 10^{-10}\,\text{s}$ are considered to be stable.\end{itemize}

\subsection[PDG\_HADRON\_MULTIPLICITIES]{PDG\_HADRON\_MULTIPLICITIES\,\cite{Amsler:2008zzb}:\\ Hadron multiplicities in hadronic $e^+e^-$ events}

\noindent Hadron multiplicities in hadronic $e^+e^-$ events, taken from Review of Particle Properties 2008, table 40.1, page 355.   Average hadron multiplicities per hadronic $e^+e^-$ annihilation event at $\sqrt{s} \approx {}$ 10, 29--35, 91, and 130--200 GeV. The numbers are averages from various experiments. Correlations of the systematic uncertainties were considered for the calculation of the averages.

\noindent\textsc{Beams:} $e^+$\,$e^-$ \newline
\textsc{Energies:} (5.0, 5.0), (17.5, 17.5), (45.6, 45.6), (88.5, 88.5) GeV \newline
\textsc{Experiment:} PDG (Various)\newline
\textsc{Spires ID:} \href{http://inspire-hep.net/search?p=find+key+7857373}{7857373}\newline
\textsc{Status:} VALIDATED\newline
\textsc{Authors:}
 \penalty 100
\begin{itemize}
  \item Hendrik Hoeth $\langle\,$\href{mailto:hendrik.hoeth@cern.ch}{hendrik.hoeth@cern.ch}$\,\rangle$
\end{itemize}
\textsc{Run details:}
 \penalty 100
\begin{itemize}

  \item Hadronic events in $e^+ e^-$ collisions\end{itemize}

\subsection[SLD\_2004\_S5693039]{SLD\_2004\_S5693039\,\cite{Abe:2003iy}:\\ Production of $\pi^+$, $\pi^-$, $K^+$, $K^-$, $p$ and $\bar p$ in Light ($uds$), $c$ and $b$ Jets from Z Decays}

\noindent Measurements of the differential production rates of stable charged particles in hadronic $Z^0$ decays, and of charged pions, kaons and protons identified over a wide momentum range. In addition to flavour-inclusive $Z^0$ decays, measurements are made for $Z^0$ decays into light ($u$, $d$, $s$), $c$ and $b$ primary flavors.

\noindent\textsc{Beams:} $e^+$\,$e^-$ \newline
\textsc{Energies:} (45.6, 45.6) GeV \newline
\textsc{Experiment:} SLD (SLC)\newline
\textsc{Spires ID:} \href{http://inspire-hep.net/search?p=find+key+5693039}{5693039}\newline
\textsc{Status:} VALIDATED\newline
\textsc{Authors:}
 \penalty 100
\begin{itemize}
  \item Peter Richardson $\langle\,$\href{mailto:Peter.Richardson@durham.ac.uk}{Peter.Richardson@durham.ac.uk}$\,\rangle$
\end{itemize}
\textsc{Run details:}
 \penalty 100
\begin{itemize}

  \item Hadronic Z decay events generated on the Z pole (\ensuremath{\sqrt{s}} = 91.2 GeV)\end{itemize}

\subsection[STAR\_2006\_S6500200]{STAR\_2006\_S6500200\,\cite{Adams:2006nd}:\\ Identified hadron spectra in pp at 200 GeV}

\noindent \pT distributions of charged pions and (anti)protons in pp collisions at $\sqrt{s} = 200$ GeV, measured by the STAR experiment at RHIC in non-single-diffractive minbias events.

\noindent\textsc{Beams:} $p$\,$p$ \newline
\textsc{Energies:} (100.0, 100.0) GeV \newline
\textsc{Experiment:} STAR (RHIC pp 200 GeV)\newline
\textsc{Spires ID:} \href{http://inspire-hep.net/search?p=find+key+6500200}{6500200}\newline
\textsc{Status:} VALIDATED\newline
\textsc{Authors:}
 \penalty 100
\begin{itemize}
  \item Bedanga Mohanty $\langle\,$\href{mailto:bedanga@rcf.bnl.gov}{bedanga@rcf.bnl.gov}$\,\rangle$
  \item Hendrik Hoeth $\langle\,$\href{mailto:hendrik.hoeth@cern.ch}{hendrik.hoeth@cern.ch}$\,\rangle$
\end{itemize}
\textsc{Run details:}
 \penalty 100
\begin{itemize}

  \item pp at 200 GeV\end{itemize}

\subsection[TASSO\_1990\_S2148048]{TASSO\_1990\_S2148048\,\cite{Braunschweig:1990yd}:\\ Event shapes in e+e- annihilation at 14-44 GeV}

\noindent Event shapes Thrust, Sphericity, Aplanarity at four different energies

\noindent\textsc{Beams:} $e^-$\,$e^+$ \newline
\textsc{Energies:} (7.0, 7.0), (11.0, 11.0), (17.5, 17.5), (21.9, 21.9) GeV \newline
\textsc{Experiment:} TASSO (PETRA)\newline
\textsc{Spires ID:} \href{http://inspire-hep.net/search?p=find+key+2148048}{2148048}\newline
\textsc{Status:} VALIDATED\newline
\textsc{Authors:}
 \penalty 100
\begin{itemize}
  \item Holger Schulz $\langle\,$\href{mailto:holger.schulz@physik.hu-berlin.de}{holger.schulz@physik.hu-berlin.de}$\,\rangle$
\end{itemize}
\textsc{Run details:}
 \penalty 100
\begin{itemize}

  \item $e^+ e^- \to$ jet jet (+ jets)  Kinematic cuts such as CKIN(1) in Pythia need to be set slightly below the CMS energy.\end{itemize}

\subsection[TOTEM\_2012\_I1115294]{TOTEM\_2012\_I1115294\,\cite{Aspell:2012ux}:\\ Forward dN/deta at 7 TeV}

\noindent The TOTEM experiment has measured the charged particle pseudorapidity density $dN_\text{ch}/d\eta$ in pp collisions at $\sqrt{s} = 7$\,TeV for $5.3 < |\eta| < 6.4$ in events with at least one charged particle with transverse momentum above 40 MeV/c in this pseudorapidity range.

\noindent\textsc{Beams:} $p$\,$p$ \newline
\textsc{Energies:} (3500.0, 3500.0) GeV \newline
\textsc{Experiment:} TOTEM (LHC)\newline
\textsc{Status:} VALIDATED\newline
\textsc{Authors:}
 \penalty 100
\begin{itemize}
  \item Hendrik Hoeth $\langle\,$\href{mailto:hendrik.hoeth@cern.ch}{hendrik.hoeth@cern.ch}$\,\rangle$
\end{itemize}
\textsc{Run details:}
 \penalty 100
\begin{itemize}

  \item pp QCD interactions at 900 GeV and 7 TeV.\end{itemize}

\subsection[UA1\_1990\_S2044935]{UA1\_1990\_S2044935\,\cite{Albajar:1989an}:\\ UA1 multiplicities, transverse momenta and transverse energy distributions.}

\noindent Particle multiplicities, transverse momenta and transverse energy distributions at the UA1 experiment, at energies of 200, 500 and 900 GeV (with one plot at 63 GeV for comparison).

\noindent\textsc{Beams:} $\bar{p}$\,$p$ \newline
\textsc{Energies:} (31.5, 31.5), (100.0, 100.0), (250.0, 250.0), (450.0, 450.0) GeV \newline
\textsc{Experiment:} UA1 (SPS)\newline
\textsc{Spires ID:} \href{http://inspire-hep.net/search?p=find+key+2044935}{2044935}\newline
\textsc{Status:} VALIDATED\newline
\textsc{Authors:}
 \penalty 100
\begin{itemize}
  \item Andy Buckley $\langle\,$\href{mailto:andy.buckley@cern.ch}{andy.buckley@cern.ch}$\,\rangle$
  \item Christophe Vaillant $\langle\,$\href{mailto:c.l.j.j.vaillant@durham.ac.uk}{c.l.j.j.vaillant@durham.ac.uk}$\,\rangle$
\end{itemize}
\textsc{Run details:}
 \penalty 100
\begin{itemize}

  \item QCD min bias events at sqrtS = 63, 200, 500 and 900 GeV.\end{itemize}

\subsection[UA5\_1982\_S875503]{UA5\_1982\_S875503\,\cite{Alpgard:1982zx}:\\ UA5 multiplicity and pseudorapidity distributions for $pp$ and $p\bar{p}$.}

\noindent Comparisons of multiplicity and pseudorapidity distributions for $pp$ and $p\bar{p}$ collisions at 53 GeV, based on the UA5 53~GeV runs in 1982. Data confirms the lack of significant difference between the two beams.

\noindent\textsc{Beams:} $\bar{p}$\,$p$, $p$\,$p$ \newline
\textsc{Energies:} (26.5, 26.5) GeV \newline
\textsc{Experiment:} UA5 (SPS)\newline
\textsc{Spires ID:} \href{http://inspire-hep.net/search?p=find+key+875503}{875503}\newline
\textsc{Status:} VALIDATED\newline
\textsc{Authors:}
 \penalty 100
\begin{itemize}
  \item Andy Buckley $\langle\,$\href{mailto:andy.buckley@cern.ch}{andy.buckley@cern.ch}$\,\rangle$
  \item Christophe Vaillant $\langle\,$\href{mailto:c.l.j.j.vaillant@durham.ac.uk}{c.l.j.j.vaillant@durham.ac.uk}$\,\rangle$
\end{itemize}
\textsc{Run details:}
 \penalty 100
\begin{itemize}

  \item Min bias QCD events at \ensuremath{\sqrt{s}} = 53~GeV. Run with both $pp$ and $p\bar{p}$ beams.\end{itemize}

\cleardoublepage

\part{How Rivet works}
\label{part:writinganalyses}
Hopefully by now you've run Rivet a few times and got the hang of the command
line interface and viewing the resulting analysis data files. Maybe you've got
some ideas of analyses that you would like to see in Rivet's library. If so,
then you'll need to know a little about Rivet's internal workings before you can
start coding: with any luck by the end of this section that won't seem
particularly intimidating.

The core objects in Rivet are ``projections'' and ``analyses''. Hopefully
``analyses'' isn't a surprise --- that's just the collection of routines that
will make histograms to compare with reference data, and the only things that
might differ there from experiences with HZTool\cite{Bromley:1995np} are the new histogramming system
and the fact that we've used some object orientation concepts to make life a bit
easier. The meaning of ``projections'', as applied to event analysis, will
probably be less obvious. We'll discuss them soon, but first a
semi-philosophical aside on the ``right way'' to do physics analyses on and
involving simulated data.

\section{The science and art of physically valid MC analysis}

The world of MC event generators is a wonderfully convenient one for
experimentalists: we are provided with fully exclusive events whose most complex
correlations can be explored and used to optimise analysis algorithms and some
kinds of detector correction effects. It is absolutely true that the majority of
data analyses and detector designs in modern collider physics would be very
different without MC simulation.

But it is very important to remember that it is just simulation: event
generators encode much of known physics and phenomenologically explore the
non-perturbative areas of QCD, but only unadulterated experiment can really tell
us about how the world behaves. The richness and convenience of MC simulation
can be seductive, and it is important that experimental use of MC strives to
understand and minimise systematic biases which may result from use of simulated
data, and to not ``unfold'' imperfect models when measuring the real world. The
canonical example of the latter effect is the unfolding of hadronisation (a
deeply non-perturbative and imperfectly-understood process) at the Tevatron (Run
I), based on MC models. Publishing ``measured quarks'' is not physics --- much
of the data thus published has proven of little use to either theory or
experiment in the following years. In the future we must be alert to such
temptation and avoid such gaffes --- and much more subtle ones.

These concerns on how MC can be abused in treating measured data also apply to
MC validation studies. A key observable in QCD tunings is the \pT of the \PZ
boson, which has no phase space at exactly $\pT = 0$ but a very sharp peak at
$\mathcal{O}(\unit{1-2}{\GeV})$. The exact location of this peak is mostly
sensitive to the width parameter of a nucleon ``intrinsic \pT'' in MC
generators, plus some soft initial state radiation and QED
bremstrahlung. Unfortunately, all the published Tevatron measurements of this
observable have either ``unfolded'' the QED effects to the ``\PZ \pT'' as
attached to the object in the HepMC/HEPEVT event record with a PDG ID code of
23, or have used MC data to fill regions of phase space where the detector could
not measure. Accordingly, it is very hard to make an accurate and portable MC
analysis to fit this data, without similarly delving into the event record in
search of ``the boson''. While common practice, this approach intrinsically
limits the precision of measured data to the calculational order of the
generator --- often not analytically well-defined. We can do better.

Away from this philosophical propaganda (which nevertheless we hope strikes some
chords in influential places\dots), there are also excellent pragmatic reasons
for MC analyses to avoid treating the MC ``truth'' record as genuine truth. The
key argument is portability: there is no MC generator which is the ideal choice
for all scenarios, and an essential tool for understanding sub-leading
variability in theoretical approaches to various areas of physics is to use
several generators with similar leading accuracies but different sub-leading
formalisms. While the HEPEVT record as written by HERWIG and PYTHIA has become
familiar to many, there are many ambiguities in how it is filled, from the
allowed graph structures to the particle content. Notably, the Sherpa event
generator explicitly elides Feynman diagram propagators from the event record,
perhaps driven by a desire to protect us from our baser analytical
instincts. The Herwig++ event generator takes the almost antipodal approach of
expressing different contributing Feynman diagram topologies in different ways
(\emph{not} physically meaningful!) and seamlessly integrating shower emissions
with the hard process particles. The general trend in MC simulation is to blur
the practically-induced line between the sampled matrix element and the
Markovian parton cascade, challenging many established assumptions about ``how
MC works''. In short, if you want to ``find'' the \PZ to see what its \pT or
$\eta$ spectrum looks like, many new generators may break your honed PYTHIA
code\dots or silently give systematically wrong results. The unfortunate truth
is that most of the event record is intended for generator debugging rather than
physics interpretation.

Fortunately, the situation is not altogether negative: in practice it is usually
as easy to write a highly functional MC analysis using only final state
particles and their physically meaningful on-shell decay parents. These are,
since the release of HepMC 2.5, standardised to have status codes of 1 and 2
respectively. \PZ-finding is then a matter of choosing decay lepton candidates,
windowing their invariant mass around the known \PZ mass, and choosing the best
\PZ candidate: effectively a simplified version of an experimental analysis of
the same quantity. This is a generally good heuristic for a safe MC analysis!
Note that since it's known that you will be running the analysis on signal
events, and there are no detector effects to deal with, almost all the details
that make a real analysis hard can be ignored. The one detail that is worth
including is summing momentum from photons around the charged leptons, before
mass-windowing: this physically corresponds to the indistinguishability of
collinear energy deposits in trackers and calorimeters and would be the ideal
published experimental measurement of Drell-Yan \pT for MC tuning. Note that
similar analyses for \PW bosons have the luxury over a true experiment of being
able to exactly identify the decay neutrino rather than having to mess around
with missing energy. Similarly, detailed unstable hadron (or tau) reconstruction
is unnecessary, due to the presence of these particles in the event record with
status code 2. In short, writing an effective analysis which is automatically
portable between generators is no harder than trying to decipher the variable
structures and multiple particle copies of the debugging-level event
objects. And of course Rivet provides lots of tools to do almost all the
standard fiddly bits for you, so there's no excuse!\\[\lineskip]

\noindent
Good luck, and be careful!

\section{Projections}

The name ``projection'' is meant to evoke thoughts of projection operators,
low-dimensional slices/views of high-dimensional spaces, and other things that
might appeal to physicists who view the world through quantum-tinted lenses. A
more mundane, but equally applicable, name would be ``observable calculators'',
but since that's a long name, the things they return aren't \emph{necessarily}
observable, and they all inherit from the \kbd{Projection} base class, we'll
stick to that name. It doesn't take long to get used to using the name as a
synonym for ``calculator'', without being intimidated by ideas that they might
be some sort of high-powered deep magic. 90\% of them is simple and
self-explanatory, as a peek under the bonnet of e.g. the all-important
\kbd{FinalState} projection will reveal.

Projections can be relatively simple things like event shapes (i.e. scalar,
vector or tensor quantities), or arbitrarily complex things like lossy or
selective views of the event final state. Most users will see them attached to
analyses by declarations in each analysis' initialisation, but they can also be
recursively ``nested'' inside other projections\footnote{Provided there are no
  dependency loops in the projection chains! Strictly, only acyclic graphs of
  projection dependencies are valid, but there is currently no code in Rivet
  that will attempt to verify this restriction.} (provided there are no infinite
loops in the nesting chain.) Calling a complex projection in an analysis may
actually transparently execute many projections on each event.

\subsection{Projection caching}

Aside from semantic issues of how the class design assigns the process of
analysing events, projections are important computationally because they live in
a framework which automatically stores (``caches'') their results between
events. This is a crucial feature for the long-term scalability of Rivet, as the
previous experience with HZTool was that HERA validation code ran very slowly
due to repeated calculation of the same $k_\perp$ clustering algorithm (at that
time notorious for scaling as the 3rd power of the number of particles.)

A concrete example may help in understanding how this works. Let's say we have
two analyses which have the same run conditions, i.e. incoming beam types, beam
energies, etc. Each also uses the thrust event shape measure to define a set of
basis vectors for their analysis. For each event that gets passed to Rivet,
whichever analysis gets called first will immediately (although maybe
indirectly) call a \kbd{FinalState} projection to get a list of stable, physical
particles (filtering out the intermediate and book-keeping entries in the HepMC
event record). That FS projection is then ``attached'' to the event. Next, the
first analysis will call a \kbd{Thrust} projection which internally uses the
same final state projection to define the momentum vectors used in calculating
the thrust. Once finished, the thrust projection will also be attached to the
event.

So far, projections have offered no benefits. However, when the second analysis
runs it will similarly try to apply its final state and thrust projections to
the event. Rather than repeat the calculations, Rivet's infrastructure will
detect that an equivalent calculation has already been run and will just return
references to the already-run projections. Since projections can also contain
and use other projections, this model allows some substantial computational
savings, without the analysis author even needing to be particularly aware of
what is going on.

Observant readers may have noticed a problem with all this projection caching
cleverness: what if the final states aren't defined the same way? One might
provide charged final state particles only, or the acceptances (defined in
pseudorapidity range and a IR \pT cutoff) might differ. Rivet handles this by
making each projection provide a comparison operator which is used to decide
whether the cached version is acceptable or if the calculation must be re-run
with different settings. Because projections can be nested, applying a top-level
projection to an event can spark off a cascade of comparisons, calculations and
cache accesses, making use of existing results wherever possible.

\subsection{Using projection caching}
So far this is all theory --- how does one actually use projections in Rivet?
First, you should understand that projections, while semantically stored within
each other, are actually all registered with a central \code{ProjectionHandler}
object.\footnote{As of version 1.1 onwards --- previously, they were stored as
  class members inside other \code{Projection}s and \code{Analysis} classes.}
The reason for this central registration is to ensure that all projections'
lifespans are managed in a consistent way, and to protect projection and
analysis authors from some technical subtleties in how C++ polymorphism works.

Inside the constructor of a \code{Projection} or the \code{init} method of an
\code{Analysis} class, you must
call the \code{addProjection} function. This takes two arguments, the projection
to be registered (by \code{const} reference), and a name. The name is local to
the parent object, so you need not worry about name clashes between objects. A
very important point is that the passed \code{Projection} is not the one that is
actually centrally registered --- that distinction belongs to a newly created
heap object which is created within the \code{addProjection} method by means of
the overloaded \code{Projection::clone()} method. Hence it is completely safe
--- and recommended --- to use only local (stack) objects in \code{Projection}
and \code{Analysis} constructors.

\begin{philosophy}
  At this point, if you have rightly bought into C++ ideas like super-strong
  type-safety, this proliferation of dynamic casting may worry you: the compiler
  can't possibly check if a projection of the requested name has been
  registered, nor whether the downcast to the requested concrete type is
  legal. These are very legitimate concerns!

  In truth, we'd like to have this level of extra safety! But in the past, when
  projections were held as members of \code{ProjectionApplier} classes rather
  than in the central \code{ProjectionHandler} repository, the benefits of the
  strong typing were outweighed by more serious and subtle bugs relating to
  projection lifetime and object ``slicing''. At least when the current approach
  goes wrong it will throw an unmissable \emph{runtime} error --- until it's
  fixed, of course! --- rather than silently do the wrong thing.

  Our problems here are a microcosm of the perpetual language battle between
  strict and dynamic typing, runtime versus compile time errors. In practice,
  this manifests itself as a trade-off between the benefits of static type
  safety and the inconvenience of the type-system gymnastics that it engenders.
  We take some comfort from the number of very good programs have been and are
  still written in dynamically typed, interpreted languages like Python, where
  virtually all error checking (barring first-scan parsing errors) must be done
  at runtime. By pushing \emph{some} checking to the domain of runtime errors,
  Rivet's code is (we believe) in practice safer, and certainly more clear and
  elegant. However, we believe that with runtime checking should come a culture
  of unit testing, which is not yet in place in Rivet.

  As a final thought, one reason for Rivet's internal complexity is that C++ is
  just not a very good language for this sort of thing: we are operating on the
  boundary between event generator codes, number crunching routines (including
  third party libraries like FastJet) and user routines. The former set
  unavoidably require native interfaces and benefit from static typing; the
  latter benefit from interface flexibility, fast prototyping and syntactic
  clarity. Maybe a future version of Rivet will break through the technical
  barriers to a hybrid approach and allow users to run compiled projections from
  interpreted analysis code. For now, however, we hope that our brand of
  ``slightly less safe C++'' will be a pleasant compromise.
\end{philosophy}





\section{Analyses}

\subsection{Writing a new analysis}

This section provides a recipe that can be followed to write a new analysis
using the Rivet projections.

Every analysis must inherit from \code{Rivet::Analysis} and, in addition to the
constructor, must implement a minimum of three methods.  Those methods are
\code{init()}, \code{analyze(const Rivet::Event\&)} and \code{finalize()}, which
are called once at the beginning of the analysis, once per event and once at the
end of the analysis respectively.

The new analysis should include the header for the base analysis class plus
whichever Rivet projections are to be used, and should work under the
\code{Rivet} namespace. Since analyses are hardly ever intended to be inherited
from, they are usually implemented within a single \kbd{.cc} file with no
corresponding header. The skeleton of a new analysis named \code{UserAnalysis}
that uses the \code{FinalState} projection might therefore start off looking
like this, in a file named \kbd{UserAnalysis.cc}:
\begin{snippet}
#include "Rivet/Analysis.hh"

namespace Rivet {

  class UserAnalysis : public Analysis {
  public:
    UserAnalysis() : Analysis("USERANA") { }
    void init() { ... }
    void analyze(const Event& event) { ... }
    void finalize() { ... }
  };

}
\end{snippet}

The constructor body is usually left empty, as all event loop setup is done in
the \code{init()} method: the one \emph{required} constructor feature is to make
a call to its base \code{Analysis} constructor, passing a string by which the
analysis will \emph{register} itself with the Rivet framework. This name is the
one exposed to a command-line or API user of this analysis: usually it is the
same as the class name, which for official analyses is always in upper case.

\begin{warning}
  Early versions of Rivet required the user to declare allowed beam types,
  energies, whether a cross-section is required, etc. in the analysis
  constructor via methods like \code{setBeams(...)} and
  \code{setNeedsCrossSection(...)}. This information is now \emph{much}
  preferred to be taken from the \kbd{.info} file for the analysis, and
  \emph{must} be done this way in analyses submitted for inclusion in future
  Rivet releases.
\end{warning}



The \code{init()} method for the \code{UserAnalysis} class should add to the analysis all
of the projections that will be used.  Projections can be added to an analysis
with a call to \code{addProjection(Projection, std::string)}, which takes as
argument the projection to be added and a name by which that projection can
later be referenced.  For this example the \code{FinalState} projection is to be
referenced by the string \code{"FS"} to provide access to all of the final state
particles inside a detector pseudorapidity coverage of $\pm 5.0$.  The syntax to
create and add that projection is as follows:
\begin{snippet}
init() {
  const FinalState fs(-5.0, 5.0);
  addProjection(fs, "FS");
}
\end{snippet}
A second task of the \code{init()} method is the booking of all histograms which
are later to be filled in the analysis code. Information about the histogramming
system can be found in Section~\ref{section:histogramming}.



\subsection{Utility classes}

Rivet provides quite a few object types for physics purposes, such as three- and
four-vectors, matrices and Lorentz boosts, and convenience proxy objects for
e.g. particles and jets. We now briefly summarise the most important features of
some of these objects; more complete interface descriptions can be found in the
generated Doxygen web pages on the Rivet web site, or simply by browsing the
relevant header files.

\subsubsection{\code{FourMomentum}}

The \code{FourMomentum} class is the main physics vector that you will encounter
when writing Rivet analyses. Its functionality and interface are similar to the
CLHEP \code{HepLorentzVector} with which many users will be familiar, but
without some of the historical baggage.

\paragraph{Vector components}%
The \code{FourMomentum} \code{E()}, \code{px()}, \code{py()}, \code{pz()} \&
\code{mass()} methods are (unsurprisingly) accessors for the vector's energy,
momentum components and mass. The \code{vector3()} method returns a spatial
\code{Vector3} object, i.e. the 3 spatial components of the 4-vector.

\paragraph{Useful properties}%
The \code{pT()} and \code{Et()} methods are used to calculate the transverse
momentum and transverse energy. Angular variables are accessed via the
\code{eta()}, \code{phi()} and \code{theta()} for the pseudorapidity, azimuthal
angle and polar angle respectively. More explicitly named versions of these also
exist, named \code{pseudorapidity()}, \code{azimuthalAngle()} and
\code{polarAngle()}. Finally, the true rapidity is accessed via the
\code{rapidity()} method. Many of these functions are also available as external
functions, as are algebraic functions such as \code{cross(vec3a, vec3b)}, which
is perhaps more palatable than \code{vec3a.cross(vec3b)}.

\paragraph{Distances}%
The $\eta$--$\phi$ distance between any two four-vectors (and/or three-vectors)
can be computed using a range of overloaded external functions of the type
\code{deltaR(vec1, vec2)}. Angles between such vectors can be calculated via the
similar \code{angle(vec1, vec2)} functions.

\subsubsection{\code{Particle}}
This class is a wrapper around the HepMC \code{GenParticle}
class. \code{Particle} objects are usually obtained as a vector from the
\code{particles()} method of a \code{FinalState} projection.  Rather than having
to directly use the HepMC objects, and e.g. translate HepMC four-vectors into
the Rivet equivalent, several key properties are accessed directly via the
\code{Particle} interface (and more may be added). The main methods of interest
are \code{momentum()}, which returns a \code{FourMomentum}, and \code{pdgId()},
which returns the PDG particle ID code. The PDG code can be used to access
particle properties by using functions such as \code{PID::isHadron()},
\code{PID::threeCharge()}, etc. (these are defined in
\kbd{Rivet/Tools/ParticleIDMethods.hh}.)

\subsubsection{\code{Jet}}
Jets are obtained from one of the jet accessor methods of a projection that
implements the \code{JetAlg} interface, e.g. \code{FastJets::jetsByPt()} (this
returns the jets sorted by \pT, such that the first element in the vector is the
hardest jet --- usually what you want.) The most useful methods are
\code{particles()}, \code{momenta()}, \code{momentum()} (a representative
\code{FourMomentum}), and some checks on the jet contents such as
\code{containsParticleId(pid)}, \code{containsCharm()} and
\code{containsBottom()}.

\subsubsection{Mathematical utilities}
The \kbd{Rivet/Math/MathUtils.hh} header defines a variety of mathematical
utility functions. These include testing functions such as \code{isZero(a)},
\code{fuzzyEquals(a, b)} and \code{inRange(a, low, high)}, whose purpose is
hopefully self-evident, and angular range-mapping functions such as
\code{mapAngle0To2Pi(a)}, \code{mapAngleMPiToPi(a)}, etc.

\subsection{Histogramming}
\label{section:histogramming}

Rivet's histogramming uses the AIDA interfaces, composed of abstract classes
\code{IHistogram1D}, \code{IProfile1D}, \code{IDataPointSet} etc. which are
built by a factories system. Since it's our feeling that much of the factory
infrastructure constitutes an abstraction overload, we provide histogram booking
functions as part of the \code{Analysis} class, so that in the \code{init}
method of your analysis you should book histograms with function calls like:
%
%
\begin{snippet}
  void init() {
    _h_one = bookHistogram1D(2,1,1);
    _h_two = bookProfile1D(3,1,2);
    _h_three = bookHistogram1D("d00-x00-y00", 50, 0.0, 1.0);
  }
\end{snippet}
Here the first two bookings have a rather cryptic 3-integer sequence as the
first arguments. This is the recommended scheme, as it makes use of the exported
data files from HepData, in which 1D histograms are constructed from a
combination of $x$ and $y$ axes in a dataset $d$, corresponding to names of the
form \kbd{d\val{d}-x\val{x}-y\val{y}}. This auto-booking of histograms saves you
from having to copy out reams of bin edges and values into your code, and makes
sure that any data fixes in HepData are easily propagated to Rivet. The
reference data files which are used for these booking methods are distributed
and installed with Rivet, you can find them in the
\kbd{\val{installdir}/share/Rivet} directory of your installation. The third
booking is for a histogram for which there is no such HepData entry: it uses the
usual scheme of specifying the name, number of bins and the min/max $x$-axis
limits manually.

Filling the histograms is done in the \code{MyAnalysis::analyse()}
function. Remember to specify the event weight as you fill:
\begin{snippet}
  void analyze(const Event& e) {
    [projections, cuts, etc.]
    ...
    _h_one->fill(pT, event.weight());
    _h_two->fill(pT, Nch, event.weight());
    _h_three->fill(fabs(eta), event.weight());
  }
\end{snippet}

Finally, histogram normalisations, scalings, divisions etc. are done in the
\code{MyAnalysis::\-finalize()} method. For normalisations and scalings you will
find appropriate convenience methods \code{Analysis::normalize(histo, norm)} and
\code{Analysis::scale(histo, scalefactor)}. Many analyses need to be scaled to
the generator cross-section, with the number of event weights to pass cuts being
included in the normalisation factor: for this you will have to track the
passed-cuts weight sum yourself via a member variable, but the analysis class
provides \code{Analysis::crossSection()} and \code{Analysis::sumOfWeights()}
methods to access the pre-cuts cross-section and weight sum respectively.

\subsection{Analysis metadata}

To keep the analysis source code uncluttered, and to allow for iteration of data
plot presentation without re-compilation and/or re-running, Rivet prefers that
analysis metadata is provided via separate files rather than hard-coded into the
analysis library. There are two such files: an \emph{analysis info} file, with
the suffix \kbd{.info}, and a \emph{plot styling} file, with the suffix
\kbd{.plot}.

\subsubsection{Analysis info files}

The analysis info files are in YAML format: a simple data format intended to be
cleaner and more human-readable/writeable than XML. As well as the analysis name
(which must coincide with the filename and the name provided to the
\kbd{Analysis} constructor, this file stores details of the collider,
experiment, date of the analysis, Rivet/data analysis authors and contact email
addresses, one-line and more complete descriptions of the analysis, advice on
how to run it, suggested generator-level cuts, and BibTeX keys and entries for
this user manual. It is also where the validation status of the analysis is declared:

See the standard analyses' info files for guidance on how to populate this
file. Info files are searched for in the paths known to the
\kbd{Rivet::getAnalysisInfoPaths()} function, which many be prepended to using
the \var{RIVET_INFO_PATH} environment variable: the first matching file to be
found will be used.

\subsubsection{Plot styling files}

The \kbd{.plot} files are in the header format for the \kbd{make-plots} plotting
system and are picked up and merged with the plot data by the Rivet
\kbd{compare-histos} script which produces the \kbd{make-plots} input data
files. All the analysis' plots should have a \kbd{BEGIN PLOT ... END PLOT}
section in this file, specifying the title and $x$/$y$-axis labels (the
\kbd{Title}, and \kbd{XLabel}/\kbd{YLabel} directives). In addition, you can use
this file to choose whether the $x$ and/or $y$ axes should be shown with a log
scale (\kbd{LogX}, \kbd{LogY}), to position the legend box to minimise clashes
with the data points and MC lines (\kbd{LegendXPos}, \kbd{LegendYPos}) and any
other valid \kbd{make-plots} directives including special text labels or forced
plot range boundaries. Regular expressions may be used to apply a directive to
all analysis names matching a pattern rather than having to specify the same
directive repeatedly for many plots.

See the standard analyses' plot files and the \kbd{make-plots} documentation
(e.g. on the Rivet website) for guidance on how to write these files. Plot info
files are searched for in the paths known to the
\kbd{Rivet::getAnalysisPlotPaths()} function, which many be prepended to using
the \var{RIVET_PLOT_PATH} environment variable. As usual, the first matching
file to be found will be used.

\subsection{Pluggable analyses}

Rivet's standard analyses are not actually built into the main \kbd{libRivet}
library: they are loaded dynamically at runtime as an analysis \emph{plugin
  library}. While you don't need to worry too much about the technicalities of
this, it does mean that you can similarly write analyses of your own, compile
them into a similar plugin library and run them from \kbd{rivet} without ever
having to modify any of the main Rivet sources or build system. This means that
you can write and run your own analyses with a system-installed copy of Rivet,
and not have to re-patch the main library when a newer version comes out
(although chances are you will have to recompile, since the binary interface usually
change between releases.)

To get started writing your analysis and understand the plugin system better,
you should check out the documentation in the wiki on the Rivet website:
\url{http://rivet.hepforge.org/trac/wiki/}. The standard
\kbd{rivet-mkanalysis} and \kbd{rivet-buildplugin} scripts can respectively be
used to make an analysis template with many ``boilerplate'' details filled in
(including bibliographic information from Inspire if available), and to build a
plugin library with the appropriate compiler options.

\subsubsection{Plugin paths}

To load pluggable analyses you will need to set the \var{RIVET_ANALYSIS_PATH}
environment variable: this is a standard colon-separated UNIX path, specifying
directories in which analysis plugin libraries may be found. If it is
unspecified, the Rivet loader system will assume that the only entry is the
\kbd{lib} directory in the Rivet installation area. Specifying the variable adds
new paths for searching \emph{before} the standard library area, and they will
be searched in the left-to-right order in the path variable. If analyses with
duplicate names are found, a warning message is issued and the first version to
have been found will be used. This allows you to override standard analyses
with same-named variants of your own, provided they are loaded from different
directories.

Several further environment variables are used to load analysis reference data
and metadata files:
\begin{description}
\item[\var{RIVET_REF_PATH}:] A standard colon-separated path list, whose
  elements are searched in order for reference histogram files. If the required
  file is not found in this path, Rivet will fall back to looking in the
  analysis library paths (for convenience, as it is normal for plugin analysis
  developers to put analysis library and data files in the same directory and it
  would be annoying to have to set several variables to make this work), and
  then the standard Rivet installation data directory.
\item[\var{RIVET_INFO_PATH}:] The path list searched first for analysis
  \kbd{.info} metadata files. The search fallback mechanism works as for
  \var{RIVET_REF_PATH}.
\item[\var{RIVET_PLOT_PATH}:] The path list searched first for analysis
  \kbd{.plot} presentation style files. The search fallbacks again work as for
  \var{RIVET_REF_PATH}.
\end{description}

These paths can be accessed from the API using the
\kbd{Rivet::getAnalysisLibPaths()} etc. functions, and can be searched for files
using the Rivet lookup rules via the \kbd{Rivet::find\-Analysis\-LibFile(filename)}
etc. functions. These functions are also available in the Python \kbd{rivet}
module. See the Doxygen documentation for more details.

\section{Using Rivet as a library}

You don't have to use Rivet via the provided command-line programmes: for some
applications you may want to have more direct control of how Rivet processes
events. Here are some possible reasons:
\begin{itemize}
\item You need to not waste CPU cycles and I/O resources on rendering HepMC
  events to a string representation which is immediately read back in. The FIFO
  idiom (Section~\ref{sec:fifo-idiom}) is not perfect: we use it in circumstances
  where the convenience and decoupling outweighs the CPU cost.
\item You don't want to write out histograms to file, preferring to use them as
  code objects. Perhaps for applications which want to manipulate histogram data
  periodically before the end of the run.
\item You enjoy tormenting Rivet developers who know their API is far from
  perfect, by complaining if it changes!
\item \dots and many more!
\end{itemize}

The Rivet API (application programming interface) has been designed in the hope
of very simple integration into other applications: all you have to do is create
a \code{Rivet::Analysis\-Handler} object, tell it which analyses to apply on the
events, and then call its \code{analyse(evt)} method for each HepMC event --
wherever they come from. The API is (we hope) stable, with the exception of the
histogramming parts.

\begin{warning}
  The histogramming interfaces in Rivet have long been advertised as marked for
  replacement, and while progress in that area has lagger far behind our
  ambitions, it \emph{will} happen with the 2.0.0 release, with unavoidable
  impact on the related parts of the API. You have been warned!
\end{warning}

The API is available for C++ and, in a more restricted form, Python. We will
explain the C++ version here; if you wish to operate Rivet (or e.g. use its
path-searching capabilities to find Rivet-related files in the standard way)
from Python then take a look inside the \kbd{rivet} and \kbd{rivet-*} Python
scripts (e.g. \kbd{less `which rivet`}) or use the module documentation cf.
\begin{snippet}
> python
>>> import rivet
>>> help(rivet)
\end{snippet}

And now the C++ API. The best way to explain is, of course, by example. Here is
a simple C++ example based on the \kbd{test/testApi.cc} source which we use in
development to ensure continuing API functionality:
\begin{snippet}
#include "Rivet/AnalysisHandler.hh"
#include "HepMC/GenEvent.h"
#include "HepMC/IO_GenEvent.h"

using namespace std;

int main() {

  // Create analysis handler
  Rivet::AnalysisHandler rivet;

  // Specify the analyses to be used
  rivet.addAnalysis("D0_2008_S7554427");
  vector<string> moreanalyses(1, "D0_2007_S7075677");
  rivet.addAnalyses(moreanalyses);

  // The usual mess of reading from a HepMC file!
  std::istream* file = new std::fstream("testApi.hepmc", std::ios::in);
  HepMC::IO_GenEvent hepmcio(*file);
  HepMC::GenEvent* evt = hepmcio.read_next_event();
  double sum_of_weights = 0.0;
  while (evt) {
    // Analyse the current event
    rivet.analyze(*evt);
    sum_of_weights += evt->weights()[0];

    // Clean up and get next event
    delete evt; evt = 0;
    hepmcio >> evt;
  }
  delete file; file = 0;

  rivet.setCrossSection(1.0);
  rivet.setSumOfWeights(sum_of_weights); // not necessary, but allowed
  rivet.finalize();
  rivet.writeData("out");

  return 0;
}
\end{snippet}

Compilation of this, if placed in a file called \kbd{myrivet.cc}, into an
executable called \kbd{myrivet} is simplest and most robust with use of the
\kbd{rivet-config} script:
\begin{snippet}
g++ myrivet.cc -o myrivet `rivet-config --cppflags --ldflags --libs`
\end{snippet}
It \emph{should} just work!

If you are doing something a bit more advanced, for example using the AGILe
package's similar API to generate Fortran generator Pythia events and pass them
directly to the Rivet analysis handler, you will need to also add the various
compiler and linker flags for the extra libraries, e.g.
\begin{snippet}
g++ myrivet.cc -o myrivet \
  `rivet-config --cppflags --ldflags --libs` \
  `agile-config --cppflags --ldflags --libs`
\end{snippet}
would be needed to compile the following AGILe+Rivet code:
\goodbreak
\begin{snippet}
#include "AGILe/Loader.hh"
#include "AGILe/Generator.hh"
#include "Rivet/AnalysisHandler.hh"
#include "HepMC/GenEvent.h"
#include "HepMC/IO_GenEvent.h"

using namespace std;

int main() {
  // Have a look what generators are available
  AGILe::Loader::initialize();
  const vector<string> gens = AGILe::Loader::getAvailableGens();
  foreach (const string& gen, gens) {
    cout << gen << endl;
  }

  // Load libraries for a specific generator and instantiate it
  AGILe::Loader::loadGenLibs("Pythia6:425");
  AGILe::Generator* generator = AGILe::Loader::createGen();
  cout << "Running " << generator->getName()
       << " version " << generator->getVersion() << endl;

  // Set generator initial state for LEP
  const int particle1 = AGILe::ELECTRON;
  const int particle2 = AGILe::POSITRON;
  const double sqrts = 91;
  generator->setInitialState(particle1, energy1, sqrts/2.0, sqrts/2.0);
  generator->setSeed(14283);

  // Set some parameters
  generator->setParam("MSTP(5)", "320"); //< PYTHIA tune
  // ...

  // Set up Rivet with a LEP analysis
  Rivet::AnalysisHandler rivet;
  rivet.addAnalysis("DELPHI_1996_S3430090");

  // Run events
  const int EVTMAX = 10000;
  HepMC::GenEvent evt;
  for (int i = 0; i < EVTMAX; ++i) {
    generator->makeEvent(evt);
    rivet.analyze(evt);
  }

  // Finalize Rivet and generator
  rivet.finalize();
  rivet.writeData("out.aida");
  generator->finalize();

  return 0;
}
\end{snippet}

%
%
%

\section{Conclusions}
\label{sec:conclusions}
We have presented a users' guide for the Rivet event-generator validation
system. This manual is intended to be a
guide to using Rivet, rather than a comprehensive
reference to the application programming interface (API) of the Rivet
library. Rivet is a C++ class library, which provides the infrastructure and
calculational tools for simulation-level analyses for high energy collider
experiments, enabling physicists to
validate event generator models and tunings with minimal effort and maximum
portability. It is designed to scale effectively to large numbers of analyses
for truly global validation, by transparent use of an automated result caching
system.

In addition to an introduction to the philosophy behind the framework, we
have given examples on how to implement the user's own analysis module.
A selected list of available analyses has been given as an example of the
flexibility of the full framework.

\cleardoublepage

\part{Appendices}
\appendix

\section{Typical \kbd{agile-runmc} commands}
\label{app:agilerunmc}
\begin{itemize}
\item \textbf{Simple run: }{\kbd{agile-runmc Herwig:6510 -P~lep1.params
      --beams=LEP:91.2 \cmdbreak -n~1000} will use the Fortran Herwig 6.5.10
    generator (the \kbd{-g} option switch) to generate 1000 events (the \kbd{-n}
    switch) in LEP1 mode, i.e. $\Ppositron\Pelectron$ collisions at $\sqrt{s} =
    \unit{91.2}{\GeV}$.}

\item \textbf{Parameter changes: }{\kbd{agile-runmc Pythia6:425
      --beams=LEP:91.2 \cmdbreak -n~1000 -P~myrun.params -p~"PARJ(82)=5.27"}
    will generate 1000 events using the Fortran Pythia 6.423 generator, again
    in LEP1 mode. The \kbd{-P} switch is actually the way of specifying a
    parameters file, with one parameter per line in the format ``\val{key}
    \val{value}'': in this case, the file \kbd{lep1.params} is loaded from the
    \kbd{\val{installdir}/share/AGILe} directory, if it isn't first found in the
    current directory.  The \kbd{-p} (lower-case) switch is used to change a
    named generator parameter, here Pythia's \kbd{PARJ(82)}, which sets the
    parton shower cutoff scale. Being able to change parameters on the command
    line is useful for scanning parameter ranges from a shell loop, or rapid
    testing of parameter values without needing to write a parameters file for
    use with~\kbd{-P}.}

\item \textbf{Writing out HepMC events: }{\kbd{agile-runmc Pythia6:425
      --beams=LHC:14TeV -n~50 -o~out.hepmc -R} will generate 50 LHC events with
    Pythia. The~\kbd{-o} switch is being used here to tell \kbd{agile-runmc} to
    write the generated events to the \kbd{out.hepmc} file. This file will be a
    plain text dump of the HepMC event records in the standard HepMC format. Use
    of filename ``-'' will result in the event stream being written to standard
    output (i.e. dumping to the terminal.}
\end{itemize}

\section{Acknowledgements}
\label{app:acknowledgements}
Rivet development has been supported by a variety of sources:

\begin{itemize}
\item All authors acknowledge support from the EU MCnet research network. MCnet
  is a Marie Curie Training Network funded under Framework Programme 6
  contract MRTN-CT-2006-035606 and Framework Programme 7 contract PITN-GA-2012-315877.
\item Andy Buckley has been supported by grants from the UK Science and
  Technology Facilities Council (Special Project Grant), the Scottish
  Universities Physics Alliance (Advanced Research Fellowship), the Institute
  for Particle Physics Phenomenology (Associateship), and a CERN Scientific
  Associateship.
\item Holger Schulz and Frank Siegert acknowledge the support of the German
  Research Foundation (DFG).
\end{itemize}

We also wish to thank the CERN MCplots (\url{http://mcplots.cern.ch}) team, and
especially Anton Karneyeu, for doing the pre-release testing since the Rivet 1.5
series and pointing out all the bits that we got wrong: Rivet is a much better
system as a result!

\cleardoublepage

\bibliographystyle{elsarticle-num}
\bibliography{refs,selectedanalyses}

\end{document}